\documentclass{article}
\usepackage{graphicx}
\usepackage{geometry}   
\usepackage{natbib} % use author/date bibliographic citations
\def\apj{\textit{ApJ}} 
 
\def\aap{\textit{A\&A}}

\def\solphys{\textit{Solar Phys.}}

\begin{document}

\begin{center}
{\large\bf The formation of a penumbra as observed with the German VTT and SoHO/MDI}

\medskip
{\bf Rolf Schlichenmaier, Nazaret Bello Gonz\'alez, \& Reza Rezaei}

Kiepenheuer Institut f\"ur Sonnenphysik, Sch\"oneckstr. 6, 79104 Freiburg, Germany

email: {\tt schliche@kis.uni-freiburg.de}

\end{center}

\paragraph{Abstract:} The generation of magnetic flux in the solar interior and its transport to the outer solar atmosphere will be in the focus of solar physics research for the next decades. One key-ingredient is the process of magnetic flux emergence into the solar photosphere, and the reorganization to form the magnetic phenomena of active regions like sunspots and pores.

On July 4, 2009, we observed a region of emerging magnetic flux, in which a proto-spot without penumbra forms a penumbra within some 4.5 hours. This process is documented by multi-wavelength observations at the German VTT: (a) imaging, (b) data with high resolution and temporal cadence acquired in Fe I 617.3 nm with the 2D imaging spectropolarimter GFPI, and (c) scans with the slit based spectropolarimeter TIP in Fe I 1089.6 nm. MDI contiuum maps and magnetograms are used to follow the formation of the proto-spot, and the subsequent evolution of the entire active region.

During the formation of the penumbra, the area and the magnetic flux of the spot increases. The additional magnetic flux is supplied by the adjacent region of emerging magnetic flux: As emerging bipole separate, the poles of the spot polarity migrate towards the spot, and finally merge with it. As more and more flux is accumulated, a penumbra forms. From inversions we infer maps for the magnetic field and the Doppler velocity (being constant along the line-of-sight). We calculate the magnetic flux of the forming spot and of the bipole footpoints that merge with the proto-spot. We witness the onset of the Evershed flow and the associated enhance of the field inclination as individual penumbral filaments form. Prior to the formation of individual penumbral sectors we detect the existence of 'counter' Evershed flows. These in-flows turn into the classical radial Evershed outflows as stable penumbra segments form.

\section{Introduction}

The solar photosphere exhibits magnetic features at a large range of scales. From these, sunspots are the largest. Simulations of flux emergence have made major progress recently \citep[e.g.,][]{cheung+al2008, tortosa+moreno2009, cheung+al2010}. Yet, high-resolution spectropolarimetric observations of such formation process are rare and either lack spectropolarimetric data or high spatial resolution \citep[][]{lites+al1998, leka+skumanich1998, yang+al2003}. In this paper, we report on high resolution spectropolarimetric observations of the formation of a sunspot penumbra. This penumbra forms around a proto-spot, thereby transforming the proto-spot into a fully developed sunspot. In addition we use MDI data to learn about how the proto-spot formed and how the entire active region develops \citep{scherrer+al1995}.

We analyze our measurements taken at the German VTT and attempt to understand how a penumbra forms. It is known \citep[see e.g. the reviews of][]{solanki2003, bellot2004, bellot2010r, borrero2009, schlichenmaier2009} that a penumbra is a site of more inclined (with respect to the vertical) magnetic fields, as compared to the field inclination in umbrae and pores. A typical penumbral feature is its radial filamentation in white light images. A radially outward directed flow is present along these filaments -- the Evershed flow. From Stokes profiles we know that this predominantly horizontal flow is magnetized \citep[e.g.,][] {bellot+al2003, bellot+al2004, franz+schlichenmaier2009}. This horizontal component is accompanied by a less inclined magnetic field component. Hence, the penumbra is a very peculiar phenomenon. And its structure is still not understood, although many model attempts have been made \citep[e.g.,][]{schlichenmaier+jahn+schmidt1998b, schlichenmaier2002,
scharmer+spruit2006, 
heinemann+al2007, ryutova+al2008, rempel+al2009b}. One way to enhance our understanding is to study sunspot formation. Therefore we analyze our unique measurements as described in \citet{schlichenmaier+al2010a, schlichenmaier+al2010b} (hereafter paper I and II, respectively) 
taken at the German VTT and try to shed new light on the formation of the penumbra.

\section{Observational setup and data acquisition}

As detailed in paper I, the campaign consisted of a multi-instrument setup and involved KAOS \citep{vdluhe+al2003}. Two imaging cameras were tuned to the G-band at 430 nm and to the Ca II K line core at 393 nm, respectively. These images were speckle reconstructed using KISIP \citep{woeger+al2008}. This was enriched by two spectropolarimeters of different type: the Fabry-P\'erot system GFPI \citep{puschmann+al2006, bello+kneer2008} and the slit-based TIP \citep{collados+al2007}. With the GFPI, we scan Fe I 617.3 nm in 56 s to measure the maps of the Stokes parameters, and with TIP we observe Fe I 1089.6 nm, with an exposure time of 10 s per slit position. 

For the GFPI and TIP data we perform inversions with SIR \citep{ruiz+deltoro1992, bellot2003} to infer physical parameters that are imprinted in the Stokes profiles by the Zeeman effect and other effects of radiative transport. To minimize the degrees of freedom, we assume that the magnetic and velocity fields are constant along the line-of-sight. Thereby, we correspondingly retrieve mean values, instead of attempting to resolve variations along the line-of-sight and laterally. 

When scanning the spot with TIP, the spot image also moves in the GFPI field-of-view. We made scans spanning not more than $2''$ with TIP to assure that the spot stays within the $30''$ by $20''$ field-of-view of GFPI. Yet, between UT 11:43 and 11:59 we performed a scan with TIP covering the entire spot. With this scan we can demonstrate that the GFPI and the TIP not only measure maps of integrated Stokes profiles that look alike, but also produce physical maps that give consistent results for the two different lines. 

\begin{figure}
% \vspace*{-2.0 cm}
\begin{center}
\includegraphics*[width=13.3cm]{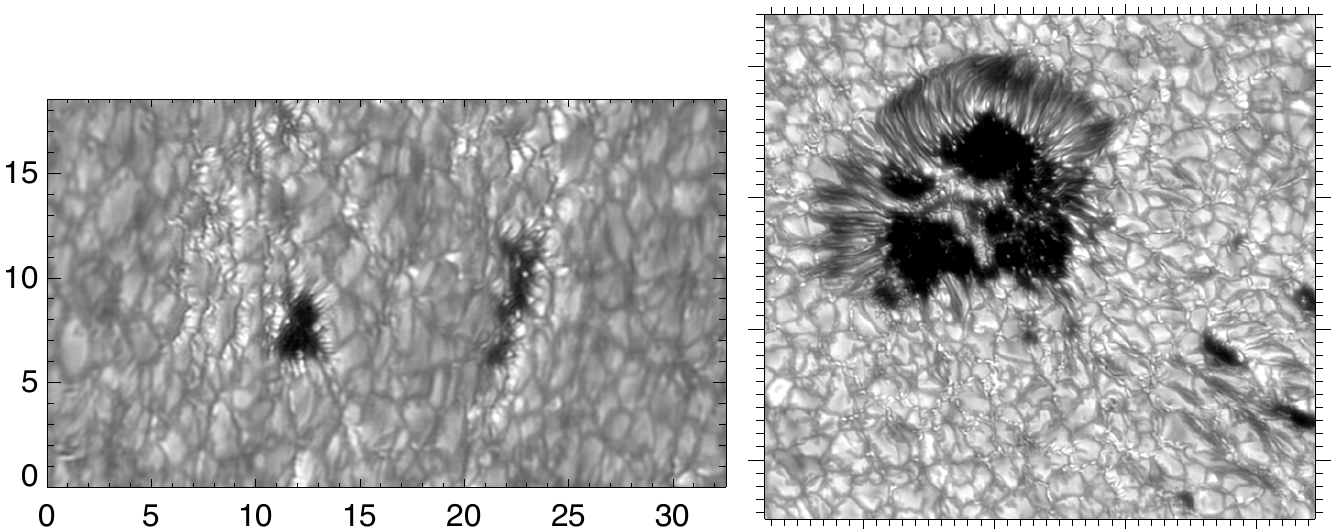}
\caption{\label{fig1} Left: On July 1st, 2009, the active region exhibits two pores at a heliocentric angle of $\theta\approx 70^\circ$ (courtesy: T.A.~Waldmann). Axis units are in arcsecs. From measurements with TIP we know, that they are of opposite polarity. During the course of the day the pores evolve and dissolve. Right: Snapshot of sunspot at $\theta =28^\circ$ on July 4 at 11:39, with tickmarks in arcsec.}
 \end{center}
\end{figure}

\begin{figure}
\vspace*{-1.0 ex}
\begin{center}
\includegraphics*{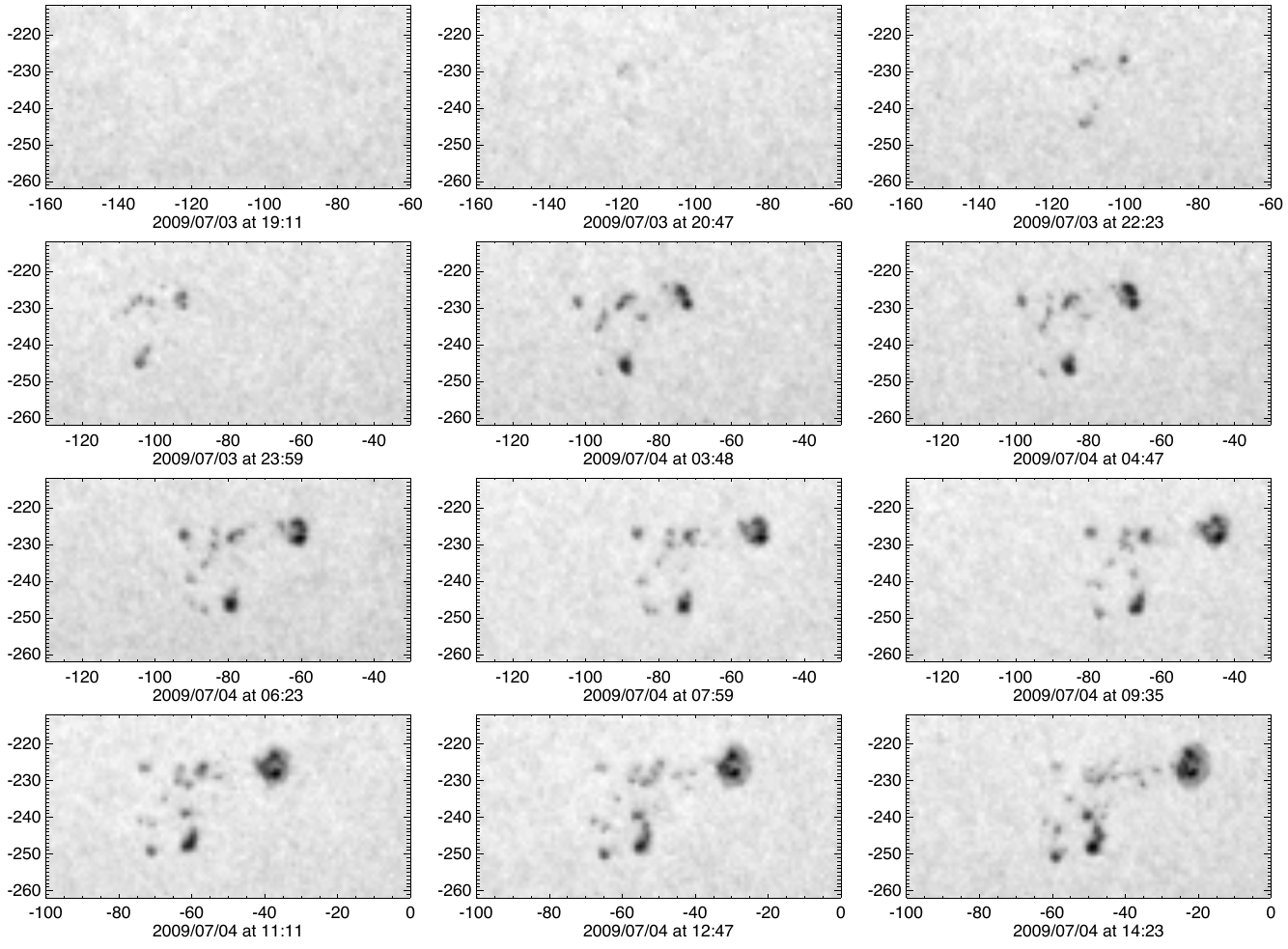}
% \vspace*{-1.0 cm}
 \caption{\label{fig2} MDI continuum images of the active region starting at 19:11 on July 3 until 14:23 on July 4. The date and time are given below each image. The axis units are in arcsec and relative to disk center. X-axis varies to keep the active region in field-of-view. }
\end{center}
\end{figure}

\begin{figure}
\begin{center}
\includegraphics*{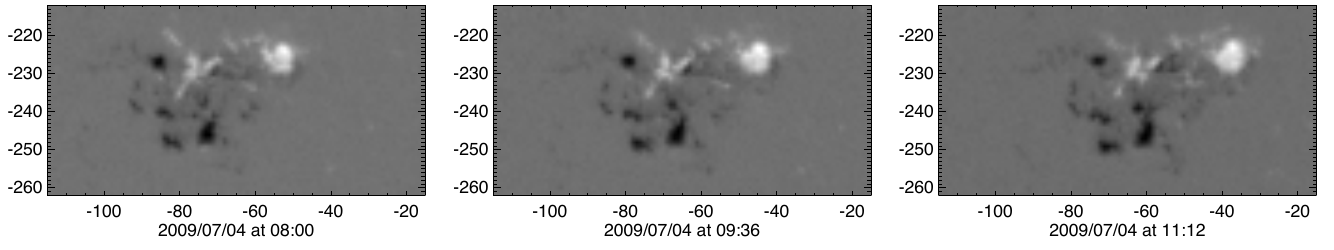} 
\caption{\label{fig3} MDI magneto-grams at 08:00, 09:36, and 11:12 on July 4. The white polarity leading spot is followed by some more diverged black polarity. Yet, the white polarity is not fully contained in the spot as there is also some white polarity flux in the emerging site between the main bulk of the polarities.}
\end{center}
\end{figure}

\section{Flux emergence and structure formation}

The spot formation on July 4, 2009, takes place in the active region NOAA 11024. This active region rotated  in onto the disk  on June 29. At this stage, it was visible as a facular region without pores. Its magnetic signatures are seen in MDI magneto-grams, but only barely in MDI continuum images. On July 1, the region formed two pores (Fig.~\ref{fig1}). These two pores -- of opposite polarity, as we know from TIP observations at the German VTT-- evolved significantly and disappeared by July 2.  In the morning of July 4, at 08:30 UT, a proto-spot with two pronounced light bridges appeared close to the zero-meridian of the Sun (Fig.~\ref{fig1}). In the following 4.5 hours a penumbra formed around the proto-spot. By 13:00 UT the seeing conditions did not allow for further observation. At that time the penumbra encircled some 220 degrees. No stable penumbra formed towards the site of magnetic flux emergence, i.e., in the direction of the opposite polarity of the active region.

Subsequently, the spot further increased in size. On July 5 the spot still showed a bridge. This light bridge disappeared by July 6. Also then, the penumbra did not fully encircle the umbra, but had a gap of 30 degrees toward the opposite polarity of the active region. The spot further evolved, exhibiting a light bridge again two days later, and rotated off the solar disk after July 10. 

\subsection{Formation of proto-spot} 

Taking advantage of the MDI data base \citep{scherrer+al1995}, we can track the formation of the proto-spot with MDI full disk continuum images. A series of maps is shown in Fig.\,\ref{fig2} spanning the time between 19:11 on July 3 until 14:23 on July 4. While no pores are visible on the first image, some dark patches are seen at 20:47 on July 3. At 22:23, there is a pronounced pore leading the active region (upper right), being located at $(-100'',-225'')$. In the first image of the second row (23:59) a second leading pore has appeared. Now the pores are at $(-92'',-230'')$, and the magneto-grams in Fig.\,\ref{fig3} indicate that they are of the same polarity. These pores are already very close to each other since their first appearance. Rather than migrating towards each other, it appears that they increase in size, and the granulation between them transforms into a light bridge. At about 07:59 (middle image of third row) they reach a state that we call proto-spot. Our observations at the VTT started at 08:30. At this stage the proto-spot started to develop penumbral segments.

The formation of the penumbra is also visible in the MDI images (bottom two rows of Fig.\,\ref{fig2}). It is seen that the penumbra forms on the side facing away from the active region. \citet{zwaan1992} -- refering to \citet{bumba+suda1984} and \citet{mcintosh1981} -- ascribes the formation of a sunspot to the coalescence of the existing pores. The case that we witness here is consistent with this picture, although the pores do not merge by approaching each other, but form very close to each other and further increase in size.

\subsection{Penumbra formation: the observational findings}

\paragraph{\bf Morphology:} Our G-band imaging and GFPI scans start around 08:30 and end at 13:05. The imaging data reveals that the penumbra forms in segments (paper I). Each segment forms in about 1 h. At the end of our time series, the penumbra does not fully spans around the umbra, but only 220 degrees (instead of 360). The segment where no stable penumbra forms is directed towards the emerging site. A snapshot of the forming sunspot at 11:39 is shown in the right panel of Fig.~\ref{fig1}.

\paragraph{\bf Elongated granules and emergent bipoles:} The emerging site is characterized by elongated granulation and intense proper motions of such granules towards and away the spot. These elongated granules play a crucial role for the penumbra formation: Such granules are associated with emerging bipoles (cf. paper II). The bipole axes, as well as the axes of their elongated granules, are oriented radially away from the center of the spot. All bipoles that we measure pop up in the photosphere such that the bipole footpoint that has the polarity of the spot is directed towards the spot. As the elongated granule increases in size, that footpoint migrates towards the spot, while the footpoint of the other polarity migrates away from the forming spot.

\paragraph{\bf Area:} The area of the spot increases from 230 arcsec$^2$ to 360 arcsec$^2$ in 4.5 hours. This increase is exclusively taken up by the formation of the penumbra, the combined area of umbra and light bridges stays constant during that time (cf., paper I).

\paragraph{\bf Magnetic flux and field strength:} From the magnetic field strength and its inclination as inferred from the inversion, we can compute the magnetic flux in the plane perpendicular to the line-of-sight, commonly referred to as the longitudinal magnetic flux. We find that this flux increases from $1.6\times 10^{21}$\,Mx to $2.4\times 10^{21}$\,Mx, i.e. the spot increases by $8\times10^{20}$\,Mx in 4 hours (08:40 until 12:38). At this rate a sunspot of $10^{22}$\,Mx would be formed in 2 days.
Since no other pore merges with the spot while the penumbra forms, we assume that all the additional flux is supplied by granular scale bipoles. We find that a typical magnetic flux value for an individual bipole element amounts to some $2$ -- $3\times 10^{18}$\,Mx. That means that about 1-2 emerging bipoles (of which the spot-polarity footpoints subsequently merge with the spot) are needed per minute to account for the increase of magnetic flux of the spot. 
As the penumbra forms, we track the magnetic field strength of the spot. We find that the mean umbral and penumbral field strength stays constant at 2.2\,kG and at 1.5\,kG, respectively. Averaging 100 pixels with the largest field strength of the umbra, we also find a constant value in time amounting to 2.7\,kG (c.f., Rezaei, Bello Gonz\'alez, \& Schlichenmaier, in preparation).

\begin{figure}
%\vspace*{-0.5ex}
\begin{center}
\includegraphics[bb=0 0 313 102]{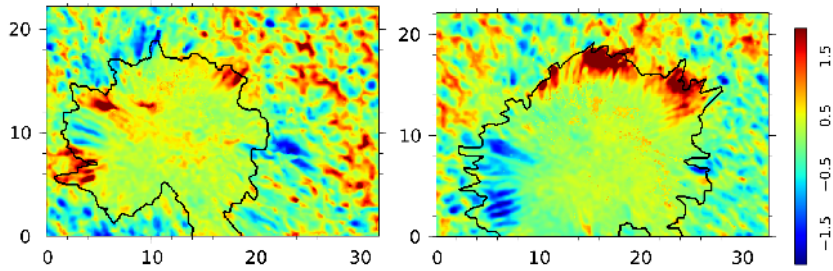}
\end{center}
\caption{\label{fig4} GFPI velocity maps inferred from inversions at 08:50 and 11:51 UT. Counter Evershed flows exist in the phase (08:50) prior to penumbra formation. Once the penumbra has formed, the flow pattern exhibits the typical Evershed flow pattern. The contours mark the white-light boundary of the spot. The color legend give the line-of-sight Doppler shift in km/s.}
\end{figure}

\paragraph{\bf Velocity field and inclination of magnetic field in penumbra:} At a heliocentric angle of $\theta=28^\circ$, the 'classical' Evershed outflow should then appear as a redshift on the limb side of the spot and as a blueshift on the center side of the spot. This is of course also true for our sunspot penumbra, but only after the penumbra has formed (cf., right panel of Fig.~\ref{fig4}). The flow pattern during the formation process is peculiar. Close inspection of flow, intensity, and inclination maps reveal that the areas of stable penumbral filaments are co-spatial with a classical Evershed flow pattern, i.e. radial outflow and large inclination. However, in areas where the penumbra has not yet formed, we observe Doppler shifts of opposite sign. These shifts can be seen in the left panel of Fig.~\ref{fig4}: Red shifts are seen in the lower left portion of the spot which is the direction toward disk center, and blue shifts are seen in the upper part which points towards the limb (i.e. should be red shifted). Most interestingly these 'counter' shifts also show a filamentary shape and are co-spatial with field inclinations of up to 80$^\circ$ (w.r.t. vertical). Hence, if these flows are along magnetic field lines, they would be radially inward, i.e., could be described as counter Evershed flows. These flows reverse their sign as the penumbra forms (c.f., Bello Gonz\'alez, Rezaei, \& Schlichenmaier, in preparation).

Presently, we can not envisage a physical scenario that could explain these counter Evershed flows. But these properties of the flow and magnetic field just before the formation of penumbral filaments certainly are most interesting and may present the key to understand how a penumbra forms.

%\smallskip 
\paragraph{Acknowledgements:}
The German VTT is operated by the Kiepenheuer-Institut f\"ur Sonnenphysik at the Observatorio del Teide in Tenerife. We acknowledge the support by the VTT and KAOS group. NBG acknowleges the Pakt f\"ur Forschung, and RR the DFG grant Schm 1168/8-2. RS is grateful for travel support from the DAAD to attend the meeting. SOHO is a project of international cooperation between ESA and NASA.

%\begin{thebibliography}{}

%\bibliographystyle{\bibpath{aa}}                           % style aa.bst 
%\bibliography{\bibpath{bibarchive_schliche}}       % your references Yourfile.bib

%\end{thebibliography}

%\begin{discussion}
%\discuss{Massey}{?}
%\discuss{van der Hucht}{!}
%\end{discussion}

\end{document}